\documentclass[english,aps,amsmath,superscriptaddress]{revtex4}

\usepackage{graphicx}
\usepackage{amsmath,amssymb,amsfonts,color}
\usepackage{wasysym}
\usepackage{ifsym}

\begin{document}

\title{Synchronization of unidirectional time delay chaotic networks and the greatest common divisor}

\author{I. Kanter}
\affiliation{Department of Physics, Bar-Ilan University, 52900 Ramat-Gan, Israel}
\author{M. Zigzag}
\affiliation{Department of Physics, Bar-Ilan University, 52900 Ramat-Gan, Israel}
\author{A. Englert}
\affiliation{Institute for Theoretical Physics, University of Wuerzburg, Am Hubland, 97074 Wuerzburg, Germany}
\author{F. Geissler}
\affiliation{Institute for Theoretical Physics, University of Wuerzburg, Am Hubland, 97074 Wuerzburg, Germany}
\author{W. Kinzel}
\affiliation{Institute for Theoretical Physics, University of Wuerzburg, Am Hubland, 97074 Wuerzburg, Germany}


\begin{abstract}
We present the interplay between synchronization of unidirectional coupled chaotic nodes with heterogeneous delays and the greatest common divisor (GCD) of loops composing the oriented graph. In the weak chaos region and for GCD=1 the network is in chaotic zero-lag synchronization, whereas for GCD=$m>1$ synchronization of m-sublattices emerges. Complete synchronization can be achieved when all chaotic nodes are influenced by an identical set of delays and in particular for the limiting case of homogeneous delays. Results are supported by simulations of chaotic systems, self-consistent and mixing arguments, as well as analytical  solutions of Bernoulli maps.
\end{abstract}

\maketitle

\section{Introduction}
Synchronization, complex networks, and chaotic dynamics with delay couplings are emerging phenomena, and concepts which have fascinated scientists for decades. These phenomena are ubiquitous in nature and play a key role in almost all fields of science including biology, ecology, physics, climatology, sociology and technology \cite{1,2,3}.  However, each one originates and is governed by different features and rules. For instance, the description and classification of complex networks are often based on their statistical properties, such as degree distribution, average degree and degree correlations \cite{3,4,5}. The observation that real networks have degree distributions that are very different from those of classical random graphs was the starting point for the recent explosion of interest in complex networks \cite{3,6}. By contrast the dynamics of processes defined on networks are closely related to the spectrum of an appropriate connection operator; a prototypical example is chaos synchronization \cite{7}, which crucially depends on the extreme eigenvalues of the graph Laplacian \cite{8,9,10}. Is there an interplay between the statistical properties of a network and its extreme spectral properties? Over the last few years a number of  papers have reported correlations between the synchronization of a network and its degree of homogeneity \cite{11,12,13}, clustering coefficients \cite{14}, degree correlations \cite{15}, average degree and degree distribution \cite{16}. The literature even reports some conflicting trends, e.g. synchronization is amplified/damaged by increasing the degree of homogeneity, or adding a few shortcut links enhances/reduces the level of synchrony \cite{11,15,16}.
The effects of distributed time delays and coupling strengths in neurons and coupled oscillators have been studied, showing that signal transmission can be seriously influenced by the distributed time delays and significant delays may result in synchronization\cite{ptp81,pre80,pre77,sen2010}. However, a general theory which relates  the structure of the network to its synchronization properties, especially in the case of networks with heterogeneous time-delay couplings and where the dynamics of the network is chaotic remains elusive.

In this Letter we report that  synchronization of chaotic networks composed of identical nonlinear units with heterogeneous unidirectional time-delayed couplings is governed by the GCD of the length of directed loops of the corresponding graphs.  We consider strongly connected oriented graphs, i.e. cases where there is a path from each node to every other node in the graph.  Each directed edge of the graph corresponds to a coupling with delay time $k_i \tau$, where $k_i$  is an integer and $\tau$ is the time unit of the coupling delay. The length of a directed loop is the sum of all integers $k_i$ along the loop, and the GCD of lengths of all loops determines the dynamic properties of the network. Two main types of chaotic synchronization can be achieved as a function of the value of the GCD in the limit of weak chaos, i.e. a small positive maximal Lyapunov exponent $\lambda_{max}$. For GCD=1, zero-lag synchronization (ZLS) among all nodes of the oriented graph can be achieved, whereas for GCD=$m>1$, nodes are partitioned into m-sublattices (clusters), where all nodes belonging to a sublattice are in ZLS. This synchronization state is different from Chimera states, where a network splits only into synchronized and desynchronized sub-populations \cite{cs101,cs91}.

\begin{figure}[t]
\includegraphics[width=0.9\textwidth,height=0.36\textwidth]{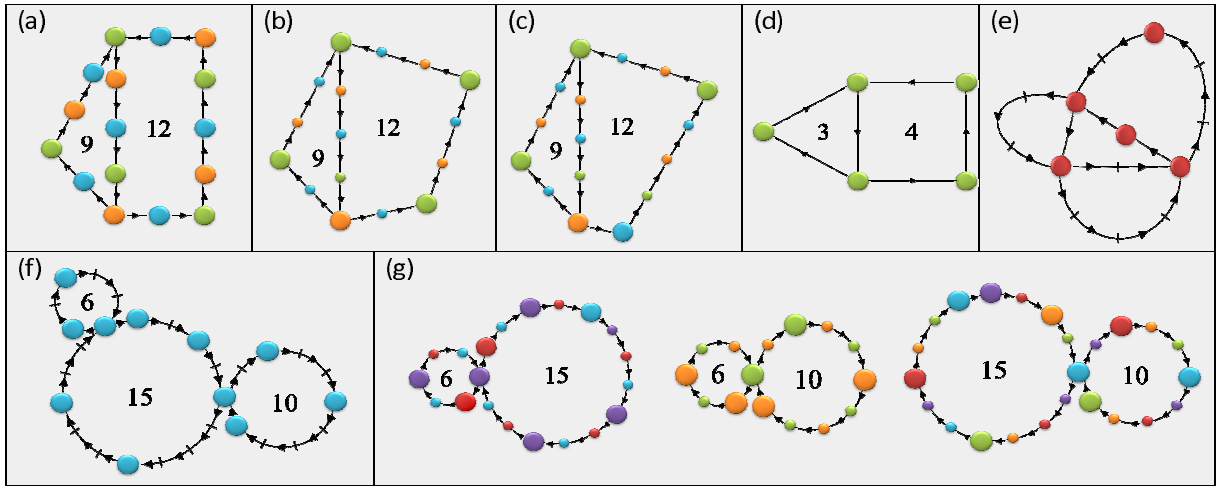}
\caption{ Synchronization of homogeneous/heterogeneous oriented chaotic motifs. Nodes with the same color are synchronized and the number of colors represents the number of SLs.  Small colored nodes in heterogeneous motifs are presented for illustration only. (a) Homogeneous motif consists of two connected loops of sizes $9\tau$ and $12\tau$ with 3-SLs. (b) Heterogeneous motif similar to (a) with 2-SLs. (c) Similar to (b) with 3-SLs. (d) Homogeneous motif similar to (a) in ZLS. (e) Heterogeneous motif in ZLS, where tick marks indicate unit distances $\tau$. (f) Three connected heterogeneous loops of sizes $6\tau$, $15\tau$ and $10\tau$, where GCD(6,10,15) =1 and the network is in ZLS. (g)  Pairs of  connected loops taken from the three loops of (f), where each pair is in SLs in contrast to ZLS in (g).
}
\end{figure}

\section{GCD-sublattices}
The results of heterogeneous oriented small networks exemplifying the role of the GCD are presented in Fig. 1 for the chaotic Bernoulli maps (BM) \cite{17}, as well as for the Lang-Kobayashi  equations (LKE) which are a good model for the intensity dynamics of coupled chaotic semiconductor lasers \cite{18,19} and are explicitly given in references \cite{19,20}.  The strength of a directed coupling is denoted by $\sigma$ and $\epsilon$ for the LKE and BM, respectively, and  for simplicity we assume that the sum of incoming equal strength couplings to each unit is identical, i.e. for the BM case and for a unit with $q$ incoming couplings the dynamics is $x_{t+1}^i=(1-\epsilon)f(x_t^i)+\epsilon/q\sum_j^q f(x_{t-k_j\tau}^j)$,  where $x_t^i$ is the state of node $i$ at time $t$ and $f(x)=(ax)mod1$ which is chaotic for $a>1$. Throughout this report $\tau$ is equal to $40$ steps and $10$ ns for BM  and  LKE , respectively,  unless otherwise  indicated.

Figure 1a depicts a network composed of two connected loops of lengths $9\tau, 12\tau$  and sixteen nodes. Since GCD(9,12)=3, a chaotic synchronization consisting of 3 sublattices (3-SLs) is expected,  and is depicted in Fig. 1a where each SL is represented by a different color. Solving the master stability function \cite{8} for the sixteen BM  reveals that for $\it{a}$=1.02, for instance, and a coupling strength  $\epsilon>0.47$ the 3-SLs is a stable solution. Figure 1b presents two loops with identical total delays as in Fig. 1a, but with only $5$ nodes, where the heterogeneous square/triangle consists of $(4\tau,\tau,4\tau,3\tau)/(4\tau,2\tau,3\tau)$ delays between connected nodes. Results of simulations for BM  with $\it{a}$=1.02 and $\epsilon>0.79$ as well as for the LKE equations with $p=1.01$ and  $\sigma=17 ns^{-1}$ indicate a 3-SLs solution, where the "color" of the $5$ nodes are fixed by the corresponding nodes in Fig. 1a. Figure 1c presents a similar heterogeneous square/triangle consisting of $(4\tau,2\tau,3\tau,3\tau)/(4\tau,2\tau,3\tau)$ delays between directed connecting nodes. Now the $5$ nodes form only 2-SLs in line with the corresponding nodes of Fig. 1a.

The main difference between homogeneous and heterogeneous oriented networks is the level of synchronization between nodes belonging to the same SL. For homogeneous networks $k_i=1$, Figs. 1a and 1d, complete synchronization is achieved such that nodes belonging to the same SL have an identical chaotic trajectory. By contrast, for heterogeneous networks,  complete synchronization is not a solution of the dynamics, since units belonging to the same SL can be connected to preceding units by different delays and chaotic signals are not periodic. Nevertheless, in the weak chaos region, a solution with a remarkable ZLS between trajectories of nodes belonging to the same SL of the heterogeneous network emerges, where the level of synchronization is enhanced toward a complete ZLS as chaos becomes weaker. For the parameters of the heterogeneous networks synchronization measured by cross correlation \cite{20} at zero time shift $C\sim0.9$, where for BM it was averaged over a window of $10^3$ steps and for  LKE  it was averaged over windows of $10$ ns, excluding the low frequency fluctuations regions \cite{19,20}.   Note that correlation below $0.5$ is the typical outcome of in vivo experiments in neuroscience, for instance, when measuring synchronization of activity among different areas of the brain \cite{21}.

The ZLS for a homogeneous network  consisting of the two connecting loops of delays  $3\tau$  and $4\tau$ (the same ratio between the two loops, $9\tau$ and $12\tau$,  of Figs. 1a-c) is presented in Fig. 1d. By rescaling  delays with the minimal delay in the network, $3\tau$, the effective relative sizes of the loops are 3 and 4 and GCD(3,4)=1.  In fact, simulations of the network with BM  as well as the analytical solution of the master stability function indicate a complete ZLS for $\it{a}$=1.02, for instance, and $\epsilon>0.12$, see Appendix. Similar results were obtained in simulations for  LKE  with $p=1.02$ and  $\sigma=13 ns^{-1}$. ZLS is also achievable for heterogeneous networks as exemplified in Fig. 1e.  The network consists of 5 nodes and loops of sizes $5,7,8,9,10$ and $12$  with GCD=1.  Simulation results indicate that $C$ is at least $0.9$ for BM with $\it{a}$=1.02 and  $\epsilon>0.89$ and for LKE  with $p=1.02$ and  $\sigma=17.5 ns^{-1}$.

A more complex heterogeneous network consisting of three connecting loops for a total delay of $6\tau (\tau,2\tau,3\tau),$ $10\tau (\tau,2\tau,3\tau,4\tau)$  and $15\tau (\tau,2\tau,2\tau,4\tau,3\tau,3\tau)$ is depicted in Fig. 1f. The GCD(6,10,15)=1 and the entire network was expected to be in ZLS, as was confirmed in simulations for  BM  with $\it{a}$=1.02 and  $\epsilon=0.92$, for instance, where $C \sim0.9$. Figure 1g depicts the synchronization of networks consisting of only a pair of connecting loops of Fig. 1f, (6,15), (6,10)  and (10,15), where the chaotic behavior is GCD(6,15)=3-SLs , GCD(6,10)=2-SLs and GCD(10,15)=5-SLs, respectively, as was confirmed in simulations for BM. These results indicate that each motif does not maintain its behavior in the entire network, which thus cannot be simply described as a "Lego" of connecting components with given chaotic modes of activity. Note that motifs can be connected either by common delay couplings or by a single node. Hence the role of GCD is a global decision which in general cannot be deduced from local topological  or geometric  properties of the network. An exception is the case where GCD=1 for two local loops, which is enough information to deduce that ZLS takes over the entire chaotic behavior of the network. Nevertheless, the GCD in general can induce long-range effects such that addition, deletion, changes in delays of existing couplings can affect the entire chaotic state of the network and in particular  correlations of remote nodes.

\section{Chain amplification}
Synchronization of  homogeneous networks  is achievable for weaker chaos, e.g. a smaller $\it{a}$ for BM, as  the size of loops increases. This trend can be attributed to the typical emergence of longer chains in a network composed of larger connected loops, where the largest Lyapunov exponent (amplification) scales in the first order approximation linearly with the number of nodes that constitute the chain, see an illustration in Fig. 2. Hence the emergence of longer chains requires a weaker chaos to maintain complete  synchronization.   A similar trend is applicable for heterogeneous networks, which resemble homogeneous networks with additional intermediate units which result in longer chains. In addition, to maintain a high level of synchronization for the heterogeneous case a weaker chaos is required than for the corresponding homogeneous network characterized by complete synchronization. Both these trends were observed in simulations and for some heterogeneous networks the  master stability functions were explicitly solved. Hence synchronization of chaotic networks is expected in general to be found in the weak chaos region.

\begin{figure}[h]
\vspace{0cm}
\hspace{0cm}
{\scalebox{1}[1]{\includegraphics[angle=-90,width=0.45\textwidth]{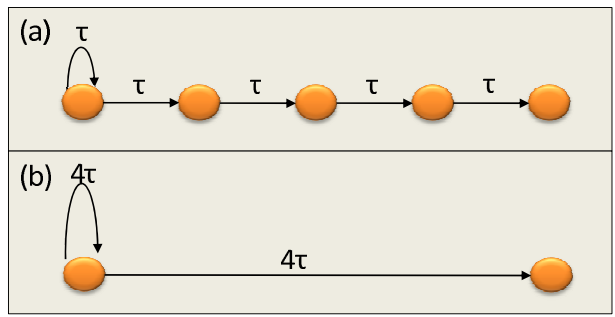}}}
\vspace{-0cm} \caption{ Two Bernoulli chains with the same length, $4\tau$: The difference between the state of the last unit is compared between two close initial conditions, $x$ and $x+\Delta$, for the first unit. For $\epsilon\rightarrow 1$, for instance, the internal dynamics is negligible and the difference for (a) is $a^4\Delta$ whereas for (b) it is $a\Delta$. Hence, the Lyapunov exponent of the entire chain is $\sim4\log(a)$ and $\sim\log(a)$ for (a) and (b), respectively. For directed graphs composed of large loops, longer chains are more likely to appear and in order to compensate the amplification of such chains, compare to graph with smaller loops, a smaller slope $a$ has to be selected. }
\end{figure}


\section{Multiple delays}
For networks with GCD=1, complete ZLS is achievable either for the case where all nodes are influenced by an identical set of heterogeneous delays or for homogeneous delays, $k_i=1$. Figure 3a depicts the heterogeneous motif with minimal number of nodes with complete ZLS. This motif consists of 3 nodes, each one of which is connected to the preceding node with two delays, $\tau$ and $2\tau$  and the motif  consists of loops of  $3\tau, 4\tau, 5\tau$ and $6\tau$ such that GCD=1. Similarly, ZLS can be found where $2\tau$  is replaced by $3\tau,~5\tau$  etc., where GCD=1. Note that the topology of this motif is similar to Fig. 1e; however, the geometry is different. Figure 3b depicts the homogeneous motif with minimal number of nodes, which consists of 4 nodes composed of loops of $3\tau$  and $4\tau$  such that GCD=1. Results of simulations for BM as well as the analytical solution of the master stability function for both networks indicate ZLS for $\it{a}$=1.02, $\epsilon>0.1$ for Fig. 3a and  $\epsilon>0.12$ for Fig. 3b. Similarly,  correlation close to one ($\sim0.99$) in the ZLS state was measured in simulations out of the low frequency fluctuations regions for LKE \cite{19,20}  for $p$=1.02 and $\sigma=14 ns^{-1}$ for Fig. 3a and  $\sigma=17.5 ns^{-1}$ for Fig. 3b. Note that complete ZLS can be found in a smaller heterogeneous motif than in homogeneous motifs and furthermore multiple delays enlarge the region, e.g. $\epsilon$ in Bernoulli, where complete ZLS is achieved.

\begin{figure}[h]
\includegraphics[width=0.6\textwidth,height=0.24\textwidth]{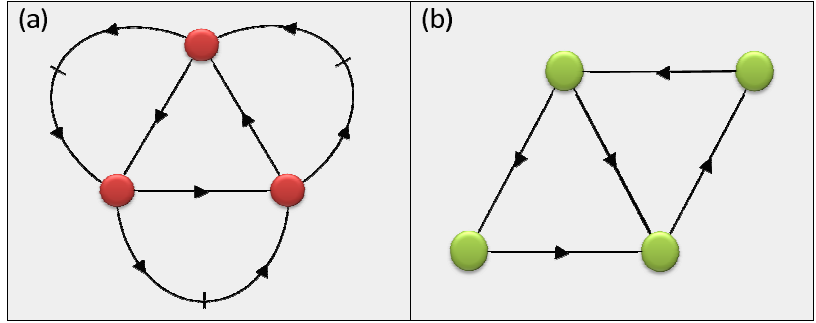}
\caption{  Oriented motifs with minimal number of units exhibit complete ZLS. (a) Heterogeneous motif consists of 3 nodes, where tick marks indicate unit distances. (b) Homogeneous motif consists of 4 nodes.}
\end{figure}

\section{Mixing argument}
The role of the GCD can be best  understood by the self-consistent argument that a necessary condition for a chaotic synchronization is that each node is driven by the same set of "colors", where for heterogeneous networks the missing nodes are artificially inserted, similar to Figs. 1b, 1c and 1g. The trivial solution is always one "color", ZLS; however, the alternative solution consists of exactly GCD "colors", GCD-SLs. An attempt to consistently color the network with a smaller number fails, since a contradiction emerges where nodes with the same color have different drives. As for bidirectional chaotic networks \cite{17}, the results always indicate that with a lack of self-couplings, SL always takes over ZLS when it is a consistent solution.  Another interesting argument that accounts for the emergence of ZLS in oriented graphs is the mixing argument which was first proposed for bidirectional networks with multiple delays \cite{23} and was recently observed in an experiment on mutually coupled chaotic lasers \cite{24}.
Figure 4a depicts the adjacency matrix, G, of Fig. 3b. The $40th$ power of the matrix $G$ indicates that it is a primitive matrix \cite{25}, where each one of the four nodes receives at time $t$ an input from all four nodes, including the node itself, from time step $t-40$, where time steps are normalized with $\tau$. Furthermore the drives for all nodes are identical indicating that only ZLS is a consistent solution. Figure 4b depicts a homogeneous graph consists of two connected loops of lengths $3\tau$ and $6\tau$ with 3-SLs and its adjacency matrix. The  $40th$ power of the matrix $G$ indicates 3-SLs: rows/colunms (1,4,7) are identical as well as rows/columns (3,5) and (2,6). All nodes belonging to the same clusters are mixing the same information from $t-40$. The structure of the matrix, $G^{40}$, also indicates that for instance the cluster (1,4,7) is driven by the cluster (3,5).

\begin{figure}[h]
\vspace{1cm}
{\scalebox{1}[1]{\includegraphics[angle=270,width=0.6\textwidth]{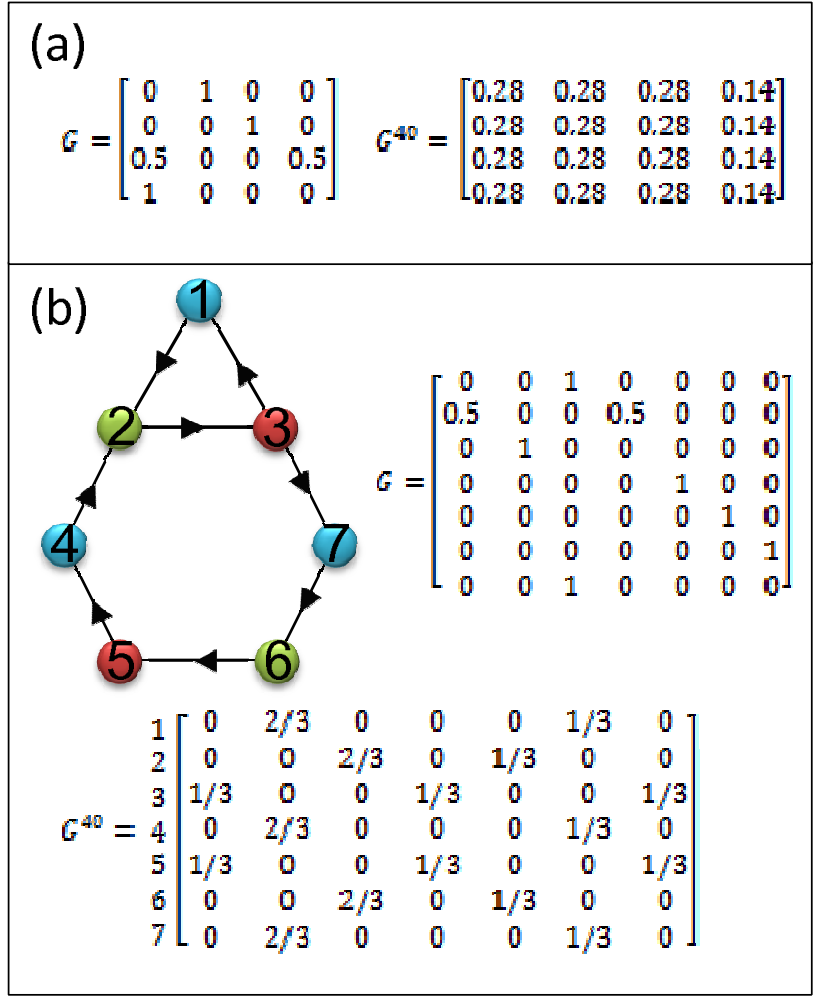}}}
\caption{  (a) The adjacency matrix, G, for the network of Fig. 3b and its $40th$ power (first two leading digits of each matrix element). (b) Two connected loops of  $3\tau$  and $6\tau$ with 3-SLs. The adjacency matrix, G, and its $40th$ power.}
\end{figure}

\begin{figure}[h]
\includegraphics[width=0.66\textwidth,height=0.24\textwidth]{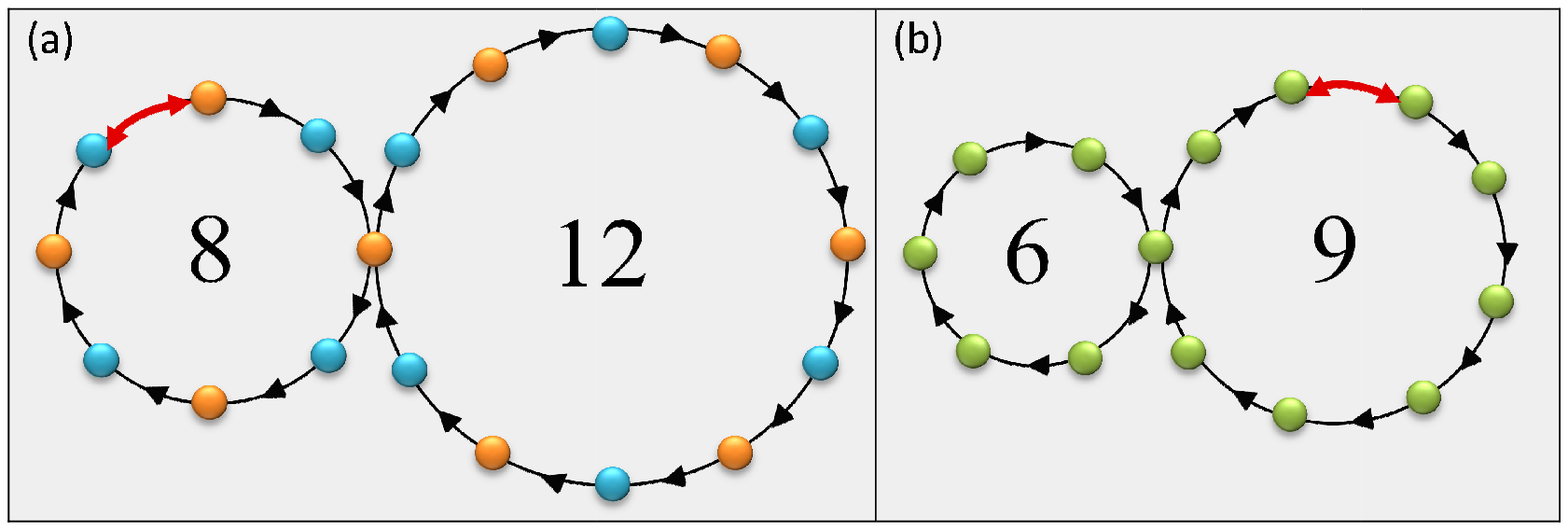}
\caption{  Change of one directed coupling to bidirectional. (a) Two connected loops   of  $8\tau$  and $12\tau$, where one bidirectional coupling (red) changes synchronization from 4-SLs to 2-SLs. (b) Two connected loops  of $6\tau$  and $9\tau$ , where one bidirectional coupling (red) changes the synchronization from 3-SLs to ZLS.}
\end{figure}

A change in only one directed coupling to bidirectional (mutual) has a dramatic effect on the synchronization pattern of a heterogeneous/homogeneous network, since a loop of size 2 is now embedded in the network.  As a result, in the case where 2 is a common divisor of all loops of the network 2-SLs takes over, otherwise ZLS is the solution. The effect of one bidirectional coupling is exemplified in Fig. 5 for homogeneous networks, where similar results were obtained for heterogeneous networks as well. Figure 5a consists of two connected directed loops   of $8\tau$ and $12\tau$ with GCD=4. After one directed coupling is converted to bidirectional (red coupling) 2-SLs is the only possible synchronization, e.g. $\it{a}$=1.01 and  $\epsilon>0.24$. Figure 5b consists of two connected loops   $6\tau$ and $9\tau$ with GCD=3, where a bidirectional coupling drives the network from 3-SLs to ZLS, e.g. $\it{a}$=1.01 and $\epsilon>0.04$. Note that the effect of a bidirectional delay, $\tau$,  is similar to the same  network with additional  self-coupling $2\tau$,  as well as the reverse operation where the loop of length $6\tau$ in Fig. 5a, for instance, can be replaced by self-coupling of $6\tau$, which were confirmed in simulations.

\section{Analytical results $\tau\rightarrow\infty$}
For homogeneous oriented Bernoulli networks in the limit of infinite delays, $\tau\rightarrow\infty$, the role of the GCD can be established analytically. The corresponding equations for the BM are
\begin{equation}
x_t^i=(1-\epsilon)f(x_{t-1}^i)+\epsilon\sum_{j\ne i} G_{ij} f(x_{t-\tau}^j )~~.
\nonumber
\end{equation}
The adjacency matrix $G_{ij}$ represents the couplings and their weights in the oriented network and we assume that the sum of the incoming signals to each unit is equal to one,  $\sum_jG_{ij}=1$. These special types of non-negative matrices are known as stochastic matrices and play a central role in Markov chain processes, where many of their mathematical properties are known \cite{25}.

For Bernoulli networks with homogeneous infinite delay couplings the master stability function depends solely on the eigenvalue spectrum of the adjacency matrix G \cite{17}. More precisely, the matrix G always has an eigenvalue  $\gamma_0=1$, which determines the Lyapunov exponent tangential to the synchronization manifold (SM) and does not affect the stability of the synchronization. The stability of the SM is given by
\begin{equation}
\gamma<e^{-\lambda_{max}\tau}
\nonumber
\end{equation}
where $\gamma$ is the second largest modulus of the eigenvalues of $G$, and  $\lambda_{max}$ is the largest Lyapunov exponent. Hence, a sufficient condition for the stability of the SM is determined by a non-zero eigenvalue gap ($\gamma<1$). Markov chain theory   now makes it possible to derive mathematical statements such as: (a) bi-partite networks as well as directed rings have $\gamma=1$ and complete ZLS is unstable; (b) a network where each node is connected to any other node and GCD=1, $\gamma<1$, hence  complete ZLS is possible; (c) for a similar network with GCD=m, $G^m$ has a block structure of m blocks and the network is in m-SLs.

\section{Concluding remarks}
The interplay between GCD and types of synchronization was found to be robust for chaotic networks with some heterogeneity of coupling strengths as well as for networks with self-couplings which is nothing else but additional directed loops.   Since the GCD is typically a global quantity, synchronization of networks cannot be simply described as a "Lego" of connecting components with given synchronization modes; hence, this casts some doubt on the importance of their statistical properties \cite{26}. Small changes in geometry and topology can dramatically alter the number of GCD-SLs, such that addition/deletion of a precise coupling can serve as a remote switching mechanism in the network. However, it should be borne in mind that the sparseness of networks is crucial for the richness of synchronization modes, since for highly dense networks the GCD is expected to be typically one.

\section{Appendix}
We exemplified calculations of the master stability for the network depicted in Fig. 1d. The dynamical equations are given by
\begin{equation}
x_{n}^{i} = (1-\epsilon)f(x_{n-1}^{i})+\epsilon\sum_j G_{ij}f(x_{n-\tau}^{j}),~~    i=1, ..., 5
\end{equation}
with the adjacency matrix
\begin{equation}
G = \begin{pmatrix}0 & 0 & 1 &0&0\\0.5 & 0&0&0.5&0\\0&1&0&0&0\\0&0&0&0&1\\0&0&1&0&0 \end{pmatrix}.
\end{equation}
For a small perturbation $\delta x_{n}^{i}$ from the trajectory $x_{n}^{i}$ , one can linearize the equations
\begin{equation}
\delta x_{n}^{i} = (1-\epsilon) a \delta x_{n-1}^{i}+\epsilon\sum_j G_{ij} a \delta x_{n-\tau}^{j},~~i=1, ..., 5
\end{equation}
and using the ansatz $\delta x_{n}^{i}=c^n \delta x_{0}^{i}$ one finds
\begin{equation}
c \delta x_{0}^{i} = (1-\epsilon) a \delta x_{0}^{i}+\epsilon\sum_j G_{ij} a c^{1-\tau}\delta x_{0}^{j},~~i=1, ..., 5.
\end{equation}
The matrix G has five eigenvalues, $\gamma_{i}$. Substituting the eigenvalues into this equation we get the following characteristic polynomial:
\begin{equation}
c ^{\tau} - (1-\epsilon) a c^{\tau - 1} - \epsilon a \gamma_{i}=0,~~~i=1, ..., 5
\end{equation}
where $c=\exp{\lambda+ i \phi}$ and $\lambda = ln |c|$ is the Lyapunov exponent. The polynomial determines the entire spectrum of the Lyapunov exponents of the system. However, only the exponents transversal to the synchronization manifold are important for the stability of the synchronization. The polynomial is of order $\tau$, therefore it has $\tau$ solutions, Lyapunov exponents, for each eigenvalue, $\gamma_{i}$. The synchronization is stable when $\gamma_{i}<0$     $\forall i$ except for those correspondents to $\gamma_1=1$ , which is parallel to the synchronization manifold. Hence checking only the maximal one for each eigenvalue is sufficient. We solved the polynomial using \textit{Matlab} for $a=1.02$ and $\tau=40$ and found that the synchronization is stable, i.e. for each of the four eigenvalues the maximal Lyapunov exponent is negative, for $\epsilon \gtrsim 0.12$.

A discussion with Fabian Wirth is acknowledged. The work of I.K. is partially support by the Israel Science Foundation.



\end{document}